\documentclass[11pt,a4paper]{article}
\usepackage[T1]{fontenc}
\usepackage[utf8]{inputenc}
\usepackage{lmodern}
\usepackage[margin=27mm]{geometry}
\usepackage{amsmath,amssymb,amsthm,bm}
\usepackage{booktabs,array}
\usepackage{microtype}
\usepackage[authoryear,round]{natbib}
\usepackage[hidelinks]{hyperref}
\usepackage{enumitem}
\usepackage{float}
\usepackage{algorithm}
\usepackage{algpseudocode}
\newfloat{algorithm}{htbp}{loa}
\floatname{algorithm}{Algorithm}
\setlist{itemsep=2pt,topsep=5pt}
\newtheorem{theorem}{Theorem}
\newtheorem{proposition}{Proposition}
\theoremstyle{remark}

\newcommand{\E}{\mathbb E}
\newcommand{\Prb}{\mathbb P}
\newcommand{\HE}{\mathcal H_{\mathcal E}}
\newcommand{\energy}{\mathcal E}

\newcommand{\ind}{\mathbf 1}

\begin{document}
{\centering
	{\Large \bf Unequal-probability distributional balance and the Horvitz-Thompson estimator}\\
	\vspace{0.5cm}
	
		Blair Robertson$^{*1}$, Chris Price$^1$, Marco Reale$^1$ and Philip Davies$^1$\\
		\vspace{0.2cm}
		$^1$School of Mathematics and Statistics, University of Canterbury, New Zealand\\
		\vspace{0.2cm}
		$^*${\it email: }blair.robertson@canterbury.ac.nz\\
}

\begin{abstract}
Auxiliary information can improve estimation precision under probability sampling, but the relationship between the study and auxiliary variables may not be known well enough to specify suitable balancing variables. Distributionally balanced designs address this by targeting agreement between sample and population auxiliary distributions. We extend energy-based distributional balance to prescribed unequal inclusion probabilities by comparing the empirical sample distribution with an inclusion-probability-weighted population distribution. Under a regularity condition relating the study values divided by their inclusion probabilities to the auxiliaries, we establish an upper bound on the design variance of the Horvitz–Thompson estimator in terms of the expected energy distance. This bound motivates an algorithm that exchanges units between randomly paired candidate samples without increasing their average energy distance. Combined with random initialization and uniform selection of a final candidate, the procedure preserves the prescribed first-order inclusion probabilities. We evaluate its auxiliary-space properties and estimation performance using generated populations and the Meuse data.
\end{abstract}

\section{Introduction}
Probability sampling provides a foundation for population inference when complete enumeration is impractical. Forest inventories, biodiversity surveys and environmental monitoring programmes use samples to estimate population characteristics while quantifying uncertainty through the sampling design. Geographical coordinates, remote-sensing measurements and other auxiliary variables are often available before the study variable is measured and can help explain variation in its values. Incorporating this information into the sampling design can improve estimation precision without increasing sample size. The challenge is to identify sample properties that support precise estimation and develop designs that promote these properties while preserving prescribed inclusion probabilities.

Balanced sampling uses auxiliary information by selecting samples for which the Horvitz--Thompson (HT) estimates of specified auxiliary totals approximately equal their known population values. The cube method \citep{deville2004} draws balanced samples with equal or unequal inclusion probabilities, and \citet{chauvet2006} developed a faster implementation. Balanced sampling is particularly effective when the study variable is well approximated by a linear combination of the balancing variables. 
The balancing variables may include transformations, interactions and category indicators to represent a range of response relationships, but they must be specified
before sampling.

Spatially balanced sampling promotes sample spread over geographical space. When nearby units have similar study values, samples that are well spread across the study area can yield more precise estimates of population characteristics than clustered samples. Generalized random-tessellation stratified sampling (GRTS) achieves spread through a randomized spatial ordering followed by systematic selection \citep{stevens2004}, whereas balanced acceptance sampling (BAS) combines a random-start Halton sequence with acceptance sampling \citep{robertson2013}. Both provide geographical coverage within a probability sampling framework.

The spreading principle also applies in auxiliary spaces, where proximity reflects similarity in observed characteristics. Distances may be defined using auxiliary variables or a suitably scaled combination of auxiliary variables and geographical coordinates. Spatially correlated Poisson sampling \citep{grafstrom2012scps}, the local pivotal method \citep{grafstrom2012}, and spatial spread sampling using weakly associated vectors \citep{jauslin2020} promote sample spread by discouraging the joint selection of nearby units. Linear assignment sampling takes a different computational approach, exchanging units between candidate samples to improve their average spread \citep{robertson2025}. These methods preserve prescribed inclusion probabilities, and their effectiveness depends on whether the chosen space captures relevant response variation.

Doubly balanced sampling combines approximate balance on auxiliary totals with spatial spread in a specified space. The local cube method applies balancing updates within neighborhoods, promoting spatial separation while approximately satisfying the balancing equations \citep{grafstrom2013}. The variables defining neighborhoods may differ from the balancing variables. This allows the design to balance on auxiliaries that explain a systematic component of the response while promoting geographical spread to account for spatial dependence in the remaining variation. Alternatively, the same variables can be used for both balancing and spreading \citep{robertson2026a}. Dynamic assignment sampling \citep{robertson2024} has also been extended to combine auxiliary balance and spatial spread in sequentially constructed samples \citep{robertson2026}. The extended design also promotes both properties in contiguous subsamples starting at the first unit in the ordering, making it useful for oversampling applications.

Balancing selected auxiliary totals and spreading a sample over a specified space do not generally ensure that the sample auxiliary distribution closely matches its population target. Differences may remain in features relevant to estimation, such as nonlinear transformations or interactions not included among the balancing variables. This motivates targeting distributional agreement when the relationship between the study and auxiliary variables is not known well enough to specify suitable balancing functions.

For this purpose, \citet{grafstrom2026a} proposed an equal-probability distributionally balanced design based on optimized circular population orderings and an alternative construction using minimum tactical configurations \citep{grafstrom2026b}. These designs aim to minimize the expected energy distance between the empirical sample and population distributions. Energy distance is a useful criterion because it captures distributional differences beyond a selected set of auxiliary totals \citep{szekely2013}. Using the connection between energy distance and reproducing kernel Hilbert spaces \citep{sejdinovic2013}, \citet{grafstrom2026a} established an estimation-error bound for an associated class of study variables under equal-probability sampling. The bound motivates minimizing expected energy distance, but lower values of this criterion do not guarantee more precise estimates for every study variable.

This paper proposes an energy-based distributionally balanced design using pairwise exchanges (DBD-PE) for sampling with prescribed unequal inclusion probabilities.
Its main contributions are:
\begin{itemize}
	\item
	We define an energy-distance criterion comparing the empirical sample distribution with an inclusion-probability-weighted population distribution. Under a regularity condition on the study values divided by their inclusion probabilities, we derive an upper bound on the design variance of the HT estimator.
	
	\item We develop a pairwise exchange algorithm that updates a collection of candidate samples, with each accepted exchange reducing their average energy distance. With the proposed random initialization and uniform selection of a final candidate, the procedure preserves the prescribed first-order inclusion probabilities.
	
	\item We demonstrate the effectiveness of DBD-PE through simulation comparisons with existing designs under equal and unequal inclusion probabilities. We examine	auxiliary-space properties and, under unequal probabilities, HT precision, the accuracy of a simple local mean variance approximation, and confidence interval coverage.
\end{itemize}

The remainder of the paper is organized as follows. Section~\ref{sec:criterion} develops the energy-distance criterion and derives an estimation-error bound for
the HT estimator. Section~\ref{sec:exchange} presents the pairwise exchange algorithm and establishes its properties. Section~\ref{sec:numerical} reports the numerical results, and Section~\ref{sec:conclusion} provides concluding remarks. Mathematical proofs appear in the Appendix.

\section{Distributional balance and estimation error}\label{sec:criterion}

Let $U=\{1,\ldots,N\}$ be a finite population of known size $N$, with fixed study values $y_i$. Our objective is to estimate the total $Y=\sum_{i\in U}y_i$. A sampling design $p$ selects a random sample $S\subseteq U$ of fixed size $n$, where $1<n\leq N$, with prescribed inclusion probabilities
\begin{equation*}
\Prb_p(i\in S)=\pi_i,\qquad
0<\pi_i\leq1,\qquad
\sum_{i\in U}\pi_i=n.
\end{equation*}
Here, $\Prb_p$ and $\E_p$ denote probability and expectation under the sampling design $p$, respectively. We estimate $Y$ using the HT estimator
\begin{equation*}
\widehat Y=\sum_{i\in S}\frac{y_i}{\pi_i},
\end{equation*}
which is design unbiased, $\E_p(\widehat Y)=Y$.

We assume that auxiliary vectors $\bm x_i\in\mathbb R^d$ are available for all population units. We use Euclidean distance in the auxiliary space, with each variable scaled to reflect its intended relative influence on the sampling design.

\subsection{Weighted empirical distributions}

Under unequal-probability sampling, units with larger inclusion probabilities tend to be overrepresented in the unweighted sample auxiliary distribution. We account for unequal selection by comparing this distribution with a population distribution weighted in proportion to the inclusion probabilities. This comparison links our energy-distance criterion directly to HT estimation through the expanded study values $y_i/\pi_i$.

Let $\delta_{\bm x}$ denote a point mass at $\bm x$. The inclusion-probability-weighted population and empirical sample distributions are
\begin{equation*}
F_\pi=\frac1n\sum_{i\in U}\pi_i\delta_{\bm x_i},\qquad
F_s=\frac1n\sum_{i\in U}a_i(s)\delta_{\bm x_i},\qquad
a_i(s)=\ind\{i\in s\},
\end{equation*}
where $s\subseteq U$ is a realized sample of size $n$ and $\ind\{\cdot\}$ is the indicator function. Both distributions have nonnegative weights summing to one. The weighted population distribution is the design expectation of the empirical sample distribution:
\[
\E_p[F_S]
=\frac1n\sum_{i\in U}\E_p[a_i(S)]\delta_{\bm x_i}
=\frac1n\sum_{i\in U}\pi_i\delta_{\bm x_i}
=F_\pi.
\]
When $\pi_i=n/N$ for every unit, $F_\pi$ reduces to the uniform population distribution, recovering the equal-probability comparison between the empirical sample and population distributions.

\subsection{Energy distance}

For a fixed sample $s$, the energy distance between $F_s$ and $F_\pi$ is
\begin{equation*}
\energy\{F_s,F_\pi\}
=2\E\|X-Z\|_2-\E\|X-X'\|_2-\E\|Z-Z'\|_2,
\end{equation*}
where $X,X'\sim F_s$ and $Z,Z'\sim F_\pi$ are mutually independent. The expectations are over draws from these distributions, with $s$ held fixed. Both distributions have finite support, so all expectations are finite. Energy distance is nonnegative and equals zero if and only if $F_s=F_\pi$ \citep{szekely2013}.

To express energy distance in terms of the population auxiliary values, we define
\begin{equation*}
d_{i\ell}=\|\bm x_i-\bm x_\ell\|_2,\qquad
\bar d_i=\frac1n\sum_{\ell\in U}\pi_\ell d_{i\ell},\qquad
c_U=\frac1{n^2}\sum_{i,\ell\in U}\pi_i\pi_\ell d_{i\ell}.
\end{equation*}
The sample criterion is then
\begin{equation}\label{eq:weighted-energy}
\energy\{F_s,F_\pi\}
=\frac2n\sum_{i\in s}\bar d_i
-\frac1{n^2}\sum_{i,\ell\in s}d_{i\ell}
-c_U.
\end{equation}
The first term is twice the average distance between the empirical sample and the inclusion-probability-weighted population. The subtracted within-sample term rewards sample spread, while $c_U$ is constant across samples. For an individual sample, minimizing the first term alone can favor units near the center of the weighted population distribution, whereas maximizing within-sample distances alone can favor extreme units.

\subsection{An estimation-error bound}\label{sec:bound}

To relate energy distance to estimation error, we impose a regularity condition on the relationship between the expanded study values $y_i/\pi_i$ and the auxiliary variables. Consider the Euclidean energy kernel
\begin{equation*}
k_{\mathcal E}(\bm x,\bm z)
=\|\bm x\|_2+\|\bm z\|_2-\|\bm x-\bm z\|_2
\end{equation*}
and let $\HE$ denote its reproducing kernel Hilbert space (RKHS). For all $i \in U$, we assume that
\begin{equation}\label{eq:response}
\frac{y_i}{\pi_i}=c+g(\bm x_i),
\qquad c\in\mathbb R,\qquad g\in\HE.
\end{equation}
The intercept $c$ allows an arbitrary level for the expanded response, since every function in $\HE$ vanishes at the origin. Its contribution $c\pi_i$ to $y_i$ is estimated exactly by the HT estimator under fixed-size sampling, so only the component $g$ contributes to the estimation error.

Assumption~\eqref{eq:response} requires units with identical auxiliary vectors to have identical expanded study values. The function $g$ satisfies the reproducing property
\begin{equation*}
g(\bm x)
=\langle g,k_{\mathcal E}(\bm x,\cdot)\rangle_{\HE},
\end{equation*}
and its norm $\|g\|_{\HE}$ controls how much the expanded response can vary between nearby auxiliary points. 

\begin{theorem}\label{thm:error}
	Under assumption~\eqref{eq:response}, every sample $s\subseteq U$ of size $n$ satisfies
	\begin{equation}\label{eq:thm1}
	|\widehat Y(s)-Y|^2
	\leq n^2\|g\|_{\HE}^2\,
	\energy\{F_s,F_\pi\}.
	\end{equation}
	Consequently, under any fixed-size sampling design $p$ with the prescribed positive inclusion probabilities,
	\begin{equation}\label{eq:thm2}
	\operatorname{Var}_p(\widehat Y)
	\leq n^2\|g\|_{\HE}^2\,
	\E_p\bigl[\energy\{F_S,F_\pi\}\bigr],
	\end{equation}
	where $\operatorname{Var}_p(\widehat Y)	=\E_p[(\widehat Y-Y)^2]$ by design unbiasedness.
\end{theorem}

Theorem~\ref{thm:error} extends the equal-probability bound in Proposition~1 of \citet{grafstrom2026a} to prescribed unequal inclusion probabilities. Their equal-probability bound is recovered by setting $\pi_i=n/N$, $c=0$ and $g=(N/n)f$, where $y_i=f(\bm x_i)$ with $f\in\HE$. For a given population, prescribed inclusion probabilities and function $g$ satisfying~\eqref{eq:response}, minimizing expected energy distance minimizes the variance bound in~\eqref{eq:thm2}, but does not necessarily minimize the design variance.

The bound is most informative when $\|g\|_{\HE}$ is small. This depends on both the response relationship and the choice of inclusion probabilities, since the condition concerns $y_i/\pi_i$. Small inclusion probabilities can make the expanded study values $y_i/\pi_i$ vary substantially between units, potentially requiring a larger RKHS norm and making the bound less informative. Choosing inclusion probabilities that reflect the scale of the study values may help, but does not by itself guarantee a small norm.

\subsection{The design criterion}

Let $\mathcal P(\bm\pi)$ denote the class of sampling designs with fixed sample size $n$ and prescribed first-order inclusion probabilities $\bm\pi=(\pi_1,\ldots,\pi_N)$. We compare these designs using their expected energy distance.

Since $\E_p[a_i(S)]=\pi_i$, the expected value of the first term in~\eqref{eq:weighted-energy} is $2c_U$. Hence,
\begin{equation*}
\E_p\bigl[\energy\{F_S,F_\pi\}\bigr]
=
c_U-\frac1{n^2}
\E_p\left[\sum_{i,\ell\in S}d_{i\ell}\right].
\end{equation*}
Thus, among designs with the prescribed inclusion probabilities, minimizing expected energy distance is equivalent to maximizing the expected sum of within-sample distances. An optimal design for this criterion satisfies
\begin{equation}\label{eq:optimal-design}
p^*\in
\underset{p\in\mathcal P(\bm\pi)}{\arg\min}\;
\E_p\bigl[\energy\{F_S,F_\pi\}\bigr].
\end{equation}

Optimizing directly over the probabilities assigned to the $\binom{N}{n}$ possible samples is generally impractical. The pairwise exchange algorithm below seeks to reduce expected energy distance while preserving the prescribed inclusion probabilities, but is not guaranteed to find a global minimum.

\section{Pairwise exchange sampling}
\label{sec:exchange}

The proposed design constructs an indexed collection of candidate samples $\mathcal S=(s_1,\ldots,s_J)$, where $J>1$ is even and each $s_j\subseteq U$ contains $n$
distinct units. Each accepted exchange reduces the average energy distance, preserves unit frequencies across candidates and keeps $n$ distinct units in each candidate. One candidate index is then selected uniformly to obtain the final sample. Together with the initialization below, this procedure preserves the prescribed first-order inclusion probabilities.

\subsection{Initialization}

Following \citet{robertson2025}, we construct the initial candidate collection $\mathcal S^{(0)}$ as follows. Randomly order the population and partition $[0,n)$
into consecutive half-open intervals of lengths $\pi_i$ in that order. Construct $J$ systematic samples with starts
\begin{equation*}
\rho_j=\rho+\frac{j-1}{J},
\qquad j=1,\ldots,J,
\qquad \rho\sim\operatorname{Unif}(0,1/J).
\end{equation*}
Candidate $s_j^{(0)}$ contains the units whose intervals contain the points $\rho_j,\rho_j+1,\ldots,\rho_j+n-1$. Since $\pi_i\leq1$, each candidate contains $n$ distinct units. The candidates are dependent because they share the population ordering and random offset. Selecting a candidate index uniformly and independently of the initialization gives a systematic start that is uniform on $[0,1)$. Thus, uniform selection from the initial collection gives the prescribed first-order inclusion
probabilities $\pi_i$.

\subsection{Objective and exchange algorithm}

We use the average energy distance across candidate samples to guide the exchanges. Writing $e(s)=\energy\{F_s,F_\pi\}$, this objective is
\begin{equation}\label{eq:obj}
\overline{\mathcal E}(\mathcal S)
=\frac1J\sum_{j=1}^J e(s_j).
\end{equation}
Each candidate is equally likely to be selected, so $\overline{\mathcal E}(\mathcal S)$ is the expected energy distance for a given collection. Averaging this quantity over the random initialization and exchanges gives the design-expected energy distance in~\eqref{eq:optimal-design}.

To update the collection, consider two candidates $s$ and $t$. Let $A=s\setminus t$ and $B=t\setminus s$. An admissible exchange swaps
$u\in A$ with $v\in B$, giving
\begin{equation*}
s'=(s\setminus\{u\})\cup\{v\},
\qquad
t'=(t\setminus\{v\})\cup\{u\}.
\end{equation*}
Both samples retain $n$ distinct units, and the combined number of occurrences of each unit is unchanged.

The combined energy distance changes by
\begin{equation}\label{eq:pair-swap-change}
\Delta_{s,t}(u,v)
=e(s')+e(t')-e(s)-e(t).
\end{equation}
The terms involving $\bar d_i$ in~\eqref{eq:weighted-energy} cancel because the combined unit counts are unchanged. Contributions from shared units also cancel, so evaluating $\Delta_{s,t}(u,v)$ requires only distances among units in $A\cup B$. The resulting exchange cost is
\begin{equation*}
\Delta_{s,t}(u,v)=\frac{2}{n^2}
\left\{
\sum_{\ell\in A\setminus\{u\}}
(d_{u\ell}-d_{v\ell})
+
\sum_{\ell\in B\setminus\{v\}}
(d_{v\ell}-d_{u\ell})
\right\}.
\end{equation*}
The corresponding change in the candidate average in~\eqref{eq:obj} is $\Delta_{s,t}(u,v)/J$. Therefore, accepting an exchange with
$\Delta_{s,t}(u,v)<0$ reduces the candidate-average energy distance, although the energy distance of one candidate may increase.

Algorithm~\ref{alg} combines initialization, pairwise exchanges and final sample selection. The exchanges are performed in rounds. In each round, we randomly permute the candidate indices and pair consecutive indices. Pairs with $|s\setminus t|=0$ (identical candidates) are skipped, as are pairs with $|s\setminus t|=1$ ($\Delta_{s,t}(u,v)=0$). For each remaining pair, we find an admissible exchange minimizing $\Delta_{s,t}(u,v)$ and accept it if the minimum is negative. At most one exchange is accepted per pair per round. After at most $R_{\max}$ rounds, one candidate is selected uniformly at random. 

\begin{algorithm}
	\caption{Pairwise exchange sampling (assumes exact arithmetic)}
	\label{alg}
	\begin{enumerate}
		\item Construct $\mathcal S^{(0)}$ as described above. Set $r=0$ and $E_0=\overline{\mathcal E}(\mathcal S^{(0)})$. If $E_0=0$, proceed to final selection.
		
		\item While $r<R_{\max}$:
		\begin{enumerate}
			\item Increment $r$. Randomly permute the candidate indices and pair consecutive indices.
			
			\item For each candidate pair $(s,t)$:
			\begin{enumerate}
				\item If $|s\setminus t|\leq1$, skip this pair.
				
				\item Find an admissible exchange minimizing $\Delta_{s,t}(u,v)$, breaking ties arbitrarily.
				
				\item Accept the exchange if $\Delta_{s,t}(u,v)<0$.
			\end{enumerate}
			
			\item Denote the updated collection by $\mathcal S^{(r)}$ and calculate $E_r=\overline{\mathcal E}(\mathcal S^{(r)})$. If $E_r=0$, stop the rounds.
		\end{enumerate}
		
		\item \textbf{Final selection:} Set $R=r$. Conditional on the final collection, draw an index $K$ uniformly from $\{1,\ldots,J\}$ and return sample $s_K^{(R)}$.
	\end{enumerate}
\end{algorithm}

The following propositions establish that optimization cannot increase the candidate-average energy distance and that the design preserves the prescribed first-order inclusion probabilities. Proofs are given in the Appendix.

\begin{proposition}\label{prop:monotonicity}
	Every accepted exchange strictly decreases $\overline{\mathcal E}(\mathcal S)$. Consequently,
	\begin{equation*}
	\overline{\mathcal E}(\mathcal S^{(R)})
	\leq
	\overline{\mathcal E}(\mathcal S^{(0)}).
	\end{equation*}
\end{proposition}

Taking expectations gives
\begin{equation*}
\E\bigl[\overline{\mathcal E}(\mathcal S^{(R)})\bigr]
\leq
\E\bigl[\overline{\mathcal E}(\mathcal S^{(0)})\bigr].
\end{equation*}
Thus, optimization cannot increase the design-expected energy distance relative to uniform selection from the initial collection. However, the procedure is not guaranteed to find a global minimum.

\begin{proposition}\label{prop:inclusion}
	Under the stated initialization and exchange procedure, uniform selection of a final candidate index gives a sample of size $n$ with first-order inclusion probabilities $\pi_i$.
\end{proposition}

The prescribed inclusion probabilities hold after averaging over the random initialization. Conditional on the final candidate collection, unit $i$ is selected with probability equal to the proportion of candidates containing it, which may differ from $\pi_i$.

\subsection{Variance estimation}\label{varSec}

We adapt the equal-probability local mean variance estimator of \cite{grafstrom2026a} to unequal-probability sampling. Define the expanded study values
\begin{equation*}
z_i=\frac{y_i}{\pi_i},\qquad i\in s.
\end{equation*}
Let $G_i\subseteq s$ contain unit $i$ and its $k-1$ nearest sampled neighbors in auxiliary space, where $2\leq k\leq n$, and define the neighborhood mean
\begin{equation*}
\bar z_i=\frac1k\sum_{j\in G_i}z_j.
\end{equation*}
We use the approximate variance estimator
\begin{equation}\label{eq:local-variance}
\widehat v_{\mathrm{LM}}(\widehat Y)
=
\frac{k}{k-1}
\sum_{i\in s}(1-\pi_i)(z_i-\bar z_i)^2.
\end{equation}

The factor $1-\pi_i$ is motivated by the variance $\pi_i(1-\pi_i)$ of a unit's inclusion indicator. The quantity $\sum_{i\in s}(1-\pi_i)z_i^2$ is an unbiased estimator of the diagonal contribution to the HT variance. We apply the same weighting to the local squared deviations in~\eqref{eq:local-variance}.

When $\pi_i=n/N$, the variance estimator reduces to
\begin{equation*}
\widehat v_{\mathrm{LM}}(\widehat Y)
=
\left(1-\frac nN\right)
\frac{N^2k}{n^2(k-1)}
\sum_{i\in s}(y_i-\bar y_i)^2,
\qquad
\bar y_i=\frac1k\sum_{j\in G_i}y_j.
\end{equation*}
This is the equal-probability local mean estimator multiplied by the finite-population correction $1-n/N$. For $k=n$, it is algebraically identical to the usual variance estimator for a total under simple random sampling without replacement.

The estimator uses only first-order inclusion probabilities and local contrasts. Neither design unbiasedness nor conservativeness is guaranteed for the proposed exchange design. We assess its accuracy and confidence-interval coverage numerically in the next section.

\section{Numerical evaluation}\label{sec:numerical}

We compare DBD-PE with simple random sampling without replacement (SRS), the random pivotal method (RPM), the local pivotal method (LPM), the local cube method (Lcube), and two distributionally balanced designs (DBD-C and DBD-TC). SRS, DBD-C and DBD-TC provide equal-probability benchmarks; RPM, LPM and Lcube also accommodate
prescribed unequal inclusion probabilities.

We use the R package \texttt{BalancedSampling} \citep{grafstrom2024} for RPM, LPM (\texttt{lpm2} version) and Lcube, and \texttt{rsamplr} \citep{prentius2026} for DBD-C and DBD-TC. The methods are implemented as follows:
\begin{itemize}
	\item RPM provides a high-entropy unequal-probability reference design.
	
	\item LPM promotes spatial balance in the space defined by all auxiliary variables.
	
	\item Lcube combines balance on auxiliary totals with spatial spread. All auxiliary variables are used for both balancing and spreading \citep{robertson2026a}. Inclusion probabilities are included as the first balancing variable to ensure fixed sample sizes.
	
	\item DBD-C uses an optimized circular population ordering to construct distributionally balanced samples.
	
	\item DBD-TC constructs distributionally balanced samples using minimum tactical configurations.
	
	\item DBD-PE is the proposed design, implemented using the \textsc{Matlab} code supplied in the supplementary material.
\end{itemize}

We assess the auxiliary space properties of each design using three measures. Distributional fit is measured by the energy distance
$\energy\{F_s,F_\pi\}$ defined in~\eqref{eq:weighted-energy}. Smaller values indicate closer agreement between the empirical sample distribution and the inclusion-probability-weighted population auxiliary distribution.

Spatial balance is measured using the Voronoi approach of \cite{stevens2004}, applied in the auxiliary space. For each population unit
$j\in U$, let
\begin{equation*}
H_j(s)=
\left\{
i\in s:
\|\bm x_j-\bm x_i\|_2
=
\min_{\ell\in s}\|\bm x_j-\bm x_\ell\|_2
\right\}
\end{equation*}
be the set of its nearest sampled units. Each population unit $j$ contributes mass $\pi_j$, which is divided equally among its nearest sampled units in $H_j(s)$. The total mass assigned to sampled unit $i\in s$ is 
\begin{equation*}
v_i(s)=
\sum_{j\in U}
\frac{\pi_j\,\ind\{i\in H_j(s)\}}{|H_j(s)|}.
\end{equation*}
The target is $v_i(s)=1$ for every sampled unit, corresponding to an even allocation of inclusion-probability mass. The spatial balance measure is
\begin{equation}\label{m1}
\mathrm{SB}(s)
=
\frac1n\sum_{i\in s}\bigl(v_i(s)-1\bigr)^2.
\end{equation}
Smaller values of $\mathrm{SB}(s)$ indicate better spatial balance.

The third measure assesses balance on auxiliary totals:
\begin{equation}\label{eq:auxiliary-balance}
\mathrm{Bal}(s)=\|\widehat{\bm X}-\bm X\|_2,
\qquad
\widehat{\bm X}=\sum_{i\in s}\frac{\bm x_i}{\pi_i},
\qquad
\bm X=\sum_{i\in U}\bm x_i.
\end{equation}
Here, $\mathrm{Bal}(s)=0$ indicates exact balance on the auxiliary totals.

\subsection{Uniform populations}

To assess whether DBD-PE retains the auxiliary space properties of existing distributionally balanced designs, we repeated a simulation from \citet{grafstrom2026a}. For each dimension $d=2,5,10,20$, we generated one population of $N=1000$ independent points from the uniform distribution on $[0,1]^d$. We considered equal-probability samples of size $n=50$.

For SRS, LPM and Lcube, we drew 10,000 independent samples from each population and calculated the mean energy distance, spatial balance (SB) and auxiliary balance (Bal).

We implemented DBD-C and DBD-TC following \citet{grafstrom2026a,grafstrom2026b}, using $10^7$ optimization iterations per run and retaining the remaining default settings. For DBD-C, we optimized one population ordering and evaluated all 1000 candidate samples. For DBD-TC, we used 50 independent runs, each producing 20 candidates. For each method, the three measures were averaged over all 1000 candidates.

For DBD-PE, we used 10 independent runs, each with $J=100$ candidates and $R_{\max}=500$ rounds. The reported measures were averaged over all 1000 final candidates.

Table~\ref{T1} reports each mean relative to its corresponding SRS mean, with values below one indicating improvement. DBD-PE closely matched DBD-TC across all measures and dimensions, indicating that it retains the auxiliary space properties of existing equal probability distributionally balanced designs. Both methods had lower reported means than LPM, Lcube and DBD-C, although their gains over SRS diminished as dimension increased, particularly for spatial balance. We next evaluate DBD-PE under prescribed unequal inclusion probabilities.

\begin{table}[tbp]
\centering
\caption{Simulation results for equal-probability samples of size $n=50$ from populations of $N=1000$ points generated uniformly on $[0,1]^d$. Mean energy distance
	($\mathcal{E}$), spatial balance (SB) and auxiliary balance (Bal) are expressed as ratios to their corresponding SRS means.}\label{T1}
\begin{tabular}{lrrrrrrrr}
\toprule
Method & $d$ &$\energy$&SB&Bal&$d$ &$\energy$&SB&Bal\\
\midrule
LPM   & 2 &0.148&0.252&0.204 & 10&0.597&0.793&0.603\\
Lcube &   &0.121&0.229&0.154 &   &0.428&0.761&0.205\\
DBD-C &   &0.095&0.171&0.089 &   &0.396&0.747&0.187\\
DBD-TC&   &0.071&0.114&0.015 &   &0.355&0.720&0.096\\
DBD-PE&   &0.069&0.113&0.015 &   &0.351&0.710&0.093\\
\addlinespace
LPM   & 5 &0.404&0.520&0.421 & 20&0.728&0.887&0.730\\
Lcube &   &0.311&0.490&0.175 &   &0.495&0.889&0.255\\
DBD-C &   &0.273&0.450&0.144 &   &0.483&0.914&0.231\\
DBD-TC&   &0.235&0.392&0.051 &   &0.441&0.883&0.142\\
DBD-PE&   &0.232&0.386&0.049 &   &0.438&0.876&0.139\\
\bottomrule
\end{tabular}
\end{table}

\subsection{Meuse population}

We assess performance under unequal-probability sampling using the Meuse data \citep{pebesma2005}. The population has $N=155$ locations in the floodplain of the Meuse River near Stein, the Netherlands, with measurements of topsoil metal concentrations and environmental variables.

We select samples of size $n=20$ with inclusion probabilities proportional to copper concentration:
\begin{equation*}
\pi_i=\frac{n c_i}{\sum_{j\in U}c_j},
\qquad i\in U,
\end{equation*}
where $c_i$ is the copper concentration for unit $i$. We compare DBD-PE with LPM, Lcube and RPM, which serves as a high-entropy reference design for unequal-probability sampling. DBD-C and DBD-TC are omitted because they require equal inclusion probabilities. 

We consider two settings, both using geographical coordinates, relative elevation and organic matter percentage as auxiliaries. Setting A adds lead and zinc concentrations, with cadmium as the response. In Setting B, lead, zinc and cadmium concentrations are separate responses.

For spreading, we center and scale the auxiliaries using their population means and standard deviations. The same standardized variables are used to calculate the three auxiliary-space measures for all methods. Lcube uses the original, unstandardized variables for its balancing constraints.

For each setting, we drew 10,000 independent samples using each of RPM, LPM and Lcube. For DBD-PE, we evaluated all 10,000 final candidates from 100 independent
runs, each with $J=100$ candidates and at most $R_{\max}=500$ rounds. Candidates within a run are dependent. However, because final selection is uniform, averaging over all candidates gives unbiased estimates of the corresponding design expectations. 

Table~\ref{tab:meuse-properties} shows that DBD-PE had the lowest reported mean energy distance, spatial balance and auxiliary balance in both settings. Its mean
energy distance was approximately 40--45\% lower than that of LPM, and its mean auxiliary balance was approximately half that of Lcube. 

For each response and design, let $\widehat Y^{(1)},\ldots,\widehat Y^{(T)}$ denote the $T=10{,}000$ HT estimates of the population total $Y$. We estimate the design variance using
\begin{equation}\label{eq:meuse-empvar}
V_{\rm emp}
=
\frac1T\sum_{t=1}^{T}
\bigl(\widehat Y^{(t)}-Y\bigr)^2.
\end{equation}
Since the HT estimator is design unbiased, $V_{\rm emp}$ is unbiased for its design variance. 

Table~\ref{tab:meuse-ht} reports empirical relative bias and relative root mean squared error (RRMSE), both expressed as percentages:
\begin{equation}\label{eq:meuse-rrmse}
\begin{aligned}
\mathrm{RBias}(\widehat Y)
&=\frac{100}{T|Y|}
\sum_{t=1}^{T}\bigl(\widehat Y^{(t)}-Y\bigr),\\
\mathrm{RRMSE}(\widehat Y)
&=\frac{100}{|Y|}\sqrt{V_{\rm emp}}.
\end{aligned}
\end{equation}
Empirical relative biases were close to zero for all methods. DBD-PE had the lowest reported RRMSE in all four comparisons, approximately 10--17\% below the lower
of the LPM and Lcube RRMSEs in each case. In Setting B, these gains held for all three responses using the same auxiliary information. These results illustrate the
potential of DBD-PE for precise HT estimation under prescribed unequal inclusion probabilities.

\begin{table}[t]
	\centering
	\caption{Mean energy distance ($\energy$), spatial balance (SB) and auxiliary balance (Bal) for the Meuse population under unequal inclusion probabilities. Smaller values indicate better performance.}
	\label{tab:meuse-properties}
	\begin{tabular}{lrrrrrr}
		\toprule
		& \multicolumn{3}{c}{Setting A}
		& \multicolumn{3}{c}{Setting B}\\
		\cmidrule(lr){2-4}\cmidrule(lr){5-7}
		Method & $\energy$ & SB & Bal
		& $\energy$ & SB & Bal\\
		\midrule
		RPM    & 0.143 & 0.366 & 75.26 & 0.113 & 0.396 & 64.12\\
		LPM    & 0.053 & 0.168 & 35.80 & 0.040 & 0.205 & 30.71\\
		Lcube  & 0.057 & 0.214 & 32.12 & 0.042 & 0.237 & 28.03\\
		DBD-PE & 0.032 & 0.151 & 15.31 & 0.022 & 0.176 & 14.31\\
		\bottomrule
	\end{tabular}
\end{table}

\begin{table}[t]
	\centering
	\caption{Empirical relative bias and relative root mean squared error (RRMSE) of the HT estimator for the Meuse population. All entries are percentages. Monte Carlo standard errors for the DBD-PE RRMSEs, calculated using the 100 independent runs, ranged from 0.028 to 0.036 percentage points.}
	\label{tab:meuse-ht}
	\begin{tabular}{lrrrrrrrr}
		\toprule
		& \multicolumn{2}{c}{Setting A}
		& \multicolumn{6}{c}{Setting B}\\
		\cmidrule(lr){2-3}\cmidrule(lr){4-9}
		& \multicolumn{2}{c}{Cadmium}
		& \multicolumn{2}{c}{Lead}
		& \multicolumn{2}{c}{Zinc}
		& \multicolumn{2}{c}{Cadmium}\\
		\cmidrule(lr){2-3}
		\cmidrule(lr){4-5}
		\cmidrule(lr){6-7}
		\cmidrule(lr){8-9}
		Method
		& Bias & RRMSE
		& Bias & RRMSE
		& Bias & RRMSE
		& Bias & RRMSE\\
		\midrule
		RPM
		&  0.07 & 10.11
		&  0.02 &  7.37
		& -0.01 &  6.37
		& -0.07 & 10.12\\
		LPM
		& -0.02 &  6.27
		&  0.00 &  5.99
		&  0.03 &  4.90
		&  0.08 &  6.73\\
		Lcube
		&  0.01 &  6.03
		&  0.04 &  5.86
		&  0.04 &  4.96
		&  0.03 &  6.71\\
		DBD-PE
		& -0.01 &  5.41
		&  0.01 &  4.96
		& -0.01 &  4.15
		&  0.01 &  5.56\\
		\bottomrule
	\end{tabular}
\end{table}

\subsection{Variance estimation}

We evaluate variance estimation for the cadmium total in Settings A and B and the lead and zinc totals in Setting B, using the same samples as in the preceding
analysis. For RPM, we use the H\'ajek--Ros\'en first-order HT variance approximation \citep[Chapter~5]{Tille2020}.

The remaining variance estimators use neighborhoods containing the sampled unit and its nearest sampled neighbors, determined using Euclidean distance in the standardized auxiliary space. For LPM, we use the spatially balanced variance estimator of \citet{grafstrom2014}, and for DBD-PE we use \eqref{eq:local-variance} with $k=2$. Both use two units per neighborhood.

For Lcube, we use the doubly balanced variance estimator of \citet{grafstrom2013}, with regression residuals based on the original, unstandardized balancing variables. Each neighborhood contains $P+1$ units, where $P$ is the number of balancing variables, including the inclusion-probability variable.

For each sample $t$, we construct the nominal 95\% confidence interval
\begin{equation*}
\widehat Y^{(t)}\pm1.96\sqrt{\widehat V^{(t)}},
\end{equation*}
where $\widehat V^{(t)}$ is its variance estimate. All variance estimates yielded valid intervals.

We assess variance calibration using $\overline{\widehat V}/V_{\rm emp}$, where $\overline{\widehat V}$ is the mean estimated variance and $V_{\rm emp}$ is defined in~\eqref{eq:meuse-empvar}. Coverage is the percentage of intervals containing $Y$. Mean interval width, in the original units, is
\begin{equation*}
\frac{3.92}{T}\sum_{t=1}^{T}\sqrt{\widehat V^{(t)}}.
\end{equation*}

Tables~\ref{tab:meuse-variance-A} and~\ref{tab:meuse-variance-B} show that the DBD-PE variance approximation was conservative on average in these examples, with coverage of 96.1--96.3\% for cadmium. It was more conservative for lead and zinc in Setting B, with coverage of 97.3\% and 96.9\%, respectively.

DBD-PE gave shorter mean intervals than RPM and LPM in all four comparisons, consistent with the precision gains in Table~\ref{tab:meuse-ht}. LPM had coverage closer to 95\% in three of the four comparisons.

For cadmium in Setting A, Lcube gave shorter mean intervals than DBD-PE, but its mean variance estimate was 17\% below $V_{\rm emp}$ and coverage was 91.2\%. For cadmium in Setting B, Lcube had coverage closer to 95\% than DBD-PE (95.7\% versus 96.1\%), but wider mean intervals.

RPM's mean variance estimates closely matched $V_{\rm emp}$, but coverage was below nominal at 93.4--93.7\%. Thus, agreement in average variance does not ensure nominal coverage.

In these examples, the local mean approximation provided relatively narrow confidence intervals for DBD-PE with above-nominal coverage. This supports its practical use, although the variation in conservativeness across responses indicates room to improve variance calibration.

\begin{table}[h]
	\centering
	\caption{Performance of variance estimators for the cadmium total in Meuse Settings A and B. Intervals have nominal 95\% coverage. Mean widths are in the original units.}
	\label{tab:meuse-variance-A}
	\begin{tabular}{lrrrrrr}
		\toprule
		& \multicolumn{3}{c}{Setting A}
		& \multicolumn{3}{c}{Setting B}\\
		\cmidrule(lr){2-4}\cmidrule(lr){5-7}
		Method
		& $\overline{\widehat V}/V_{\rm emp}$
		& Coverage (\%)
		& Mean width
		& $\overline{\widehat V}/V_{\rm emp}$
		& Coverage (\%)
		& Mean width\\
		\midrule
		RPM
		& 1.01 & 93.7 & 197
		& 1.00 & 93.7 & 196\\
		LPM
		& 1.10 & 94.3 & 125
		& 1.04 & 93.6 & 131\\
		Lcube
		& 0.83 & 91.2 & 107
		& 1.14 & 95.7 & 137\\
		DBD-PE
		& 1.23 & 96.3 & 115
		& 1.25 & 96.1 & 120\\
		\bottomrule
	\end{tabular}
\end{table}

\begin{table}[h]
	\centering
	\caption{Performance of variance estimators for the lead and zinc totals in Meuse Setting B. Intervals have nominal 95\% coverage.
		Mean widths are in the original units.}
	\label{tab:meuse-variance-B}
	\begin{tabular}{lrrrrrr}
		\toprule
		& \multicolumn{3}{c}{Lead}
		& \multicolumn{3}{c}{Zinc}\\
		\cmidrule(lr){2-4}\cmidrule(lr){5-7}
		Method
		& $\overline{\widehat V}/V_{\rm emp}$
		& Coverage (\%)
		& Mean width
		& $\overline{\widehat V}/V_{\rm emp}$
		& Coverage (\%)
		& Mean width\\
		\midrule
		RPM
		& 0.99 & 93.4 & 6789
		& 0.99 & 93.4 & 17869\\
		LPM
		& 1.31 & 95.5 & 6269
		& 1.25 & 95.5 & 15278\\
		Lcube
		& 1.65 & 97.9 & 6938
		& 1.43 & 97.3 & 16637\\
		DBD-PE
		& 1.67 & 97.3 & 5870
		& 1.54 & 96.9 & 14399\\
		\bottomrule
	\end{tabular}
\end{table}

\section{Conclusion}\label{sec:conclusion}

We have extended energy-based distributionally balanced sampling to prescribed unequal inclusion probabilities. The criterion compares the empirical sample distribution with an inclusion-probability-weighted population distribution. Under a regularity condition on the expanded study values, we establish an upper bound on the design variance of the HT estimator in terms of expected energy distance. The proposed pairwise exchange algorithm updates a collection of candidate samples, with each accepted exchange reducing their average energy distance. Together with the proposed initialization and uniform final selection, the algorithm preserves the prescribed first-order inclusion probabilities.

Under equal inclusion probabilities, DBD-PE retained the favorable auxiliary space properties of existing distributionally balanced designs, closely matching
DBD-TC on the uniform populations. For the Meuse population under unequal inclusion probabilities, DBD-PE had lower reported mean energy distance, spatial balance and auxiliary balance than RPM, LPM and Lcube. These improvements were accompanied by lower RRMSEs for all responses considered, with the same auxiliary information used for all three responses in Setting B. The variance bound provides a theoretical rationale for the energy-distance criterion, and the Meuse results offer promising initial evidence of DBD-PE's practical performance. Extending the unequal-probability evaluation beyond the single population examined here will help assess how broadly the observed gains apply.

The local mean variance approximation was conservative on average, with empirical confidence interval coverage above the nominal level. DBD-PE produced shorter mean intervals than RPM and LPM in all four comparisons. These results support the practical use of the proposed design. Further work could develop a simple variance estimator that better accounts for how units are selected together.

\appendix
\section{Appendix}

\subsubsection*{Proof of Theorem~\ref{thm:error}}

\begin{proof}
	Fix a sample $s$ of size $n$ and write
	$b_i=\{a_i(s)-\pi_i\}/n$.
	Since $\sum_{i\in U}a_i(s)=\sum_{i\in U}\pi_i=n$,
	we have $\sum_{i\in U}b_i=0$. Thus, the intercept
	in~\eqref{eq:response} cancels.
	Using assumption~\eqref{eq:response} and the
	reproducing property,
	\begin{align*}
	\frac{\widehat Y(s)-Y}{n}
	&=\sum_{i\in U}b_i\frac{y_i}{\pi_i}\\
	&=\sum_{i\in U}b_i g(\bm x_i)\\
	&=\left\langle
	g,\sum_{i\in U}b_i k_{\mathcal E}(\bm x_i,\cdot)
	\right\rangle_{\HE}.
	\end{align*}
	
	The squared norm of the second argument is
	\begin{align*}
	\left\|
	\sum_{i\in U}b_i k_{\mathcal E}(\bm x_i,\cdot)
	\right\|_{\HE}^{2}
	&=\sum_{i,\ell\in U}b_i b_\ell
	k_{\mathcal E}(\bm x_i,\bm x_\ell)\\
	&=-\sum_{i,\ell\in U}b_i b_\ell d_{i\ell},
	\end{align*}
	because the terms involving $\|\bm x_i\|_2$ and
	$\|\bm x_\ell\|_2$ vanish when $\sum_{i\in U}b_i=0$.
	Substituting $b_i=\{a_i(s)-\pi_i\}/n$ gives
	\begin{align*}
	-\sum_{i,\ell\in U}b_i b_\ell d_{i\ell}
	&=-\frac1{n^2}\sum_{i,\ell\in U}
	a_i(s)a_\ell(s)d_{i\ell}\\
	&\quad+\frac1{n^2}\sum_{i,\ell\in U}
	\{a_i(s)\pi_\ell+\pi_i a_\ell(s)\}d_{i\ell}
	-\frac1{n^2}\sum_{i,\ell\in U}
	\pi_i\pi_\ell d_{i\ell}\\
	&=\frac2n\sum_{i\in U}a_i(s)\bar d_i
	-\frac1{n^2}\sum_{i,\ell\in U}
	a_i(s)a_\ell(s)d_{i\ell}
	-c_U\\
	&=\energy\{F_s,F_\pi\}.
	\end{align*}
	The second equality uses $d_{i\ell}=d_{\ell i}$ and
	the definitions of $\bar d_i$ and $c_U$.
	Hence, the Cauchy--Schwarz inequality yields
	\[
	\left|
	\frac{\widehat Y(s)-Y}{n}
	\right|^2
	\leq
	\|g\|_{\HE}^{2}
	\energy\{F_s,F_\pi\}.
	\]
	Multiplying by $n^2$ establishes~\eqref{eq:thm1}.
	Taking expectation under the sampling design and
	using the design unbiasedness of $\widehat Y$
	gives~\eqref{eq:thm2}.
\end{proof}

\subsubsection*{Proof of Proposition~\ref{prop:monotonicity}}

\begin{proof}
	Let $\mathcal S'$ be obtained from $\mathcal S$ by an
	accepted exchange between candidates $s$ and $t$.
	Only these two candidates change, so
	\[
	\overline{\mathcal E}(\mathcal S')
	-\overline{\mathcal E}(\mathcal S)
	=\frac{\Delta_{s,t}(u,v)}{J}<0.
	\]
	If no exchange is accepted, the collection remains unchanged.
	Consequently, the candidate-average energy distance cannot
	increase during any round. Applying this argument up to the
	stopping round $R$ gives
	\[
	\overline{\mathcal E}(\mathcal S^{(R)})
	\leq
	\overline{\mathcal E}(\mathcal S^{(0)}).
	\]
	This inequality holds for every realization of the algorithm,
	whether it terminates at zero energy or after $R_{\max}$ rounds.
\end{proof}

\subsubsection*{Proof of Proposition~\ref{prop:inclusion}}

\begin{proof}
	In this proof, $\E$ denotes expectation over the
	random initialization and exchange procedure.
	
	For each unit $i\in U$, let
	\[
	\nu_i(\mathcal S)
	=\sum_{j=1}^{J}\mathbf{1}\{i\in s_j\}
	\]
	be its number of occurrences across the candidate samples.
	
	First consider the initial collection. Conditional on the
	population ordering, select an index $K$ uniformly from
	$\{1,\ldots,J\}$, independently of the random offset $\rho$.
	Since $\rho\sim\operatorname{Unif}(0,1/J)$, the selected start
	\[
	\rho_K=\rho+\frac{K-1}{J}
	\]
	is uniform on $[0,1)$.
	
	Unit $i$ corresponds to a half-open interval of length
	$0 < \pi_i\leq1$. The systematic points are spaced one unit
	apart, so at most one can fall in this interval. For a
	uniform start on $[0,1)$, the probability that one does
	is its length, $\pi_i$. Thus, uniform selection from the
	initial collection gives
	\[
	\mathbb E\left[
	\frac{\nu_i(\mathcal S^{(0)})}{J}
	\right]=\pi_i.
	\]
	
	Every admissible exchange moves each exchanged unit from
	one candidate to the other. It therefore preserves
	$\nu_i(\mathcal S)$ for every unit and leaves $n$ distinct
	units in each candidate. Hence,
	\[
	\nu_i(\mathcal S^{(R)})
	=\nu_i(\mathcal S^{(0)}),
	\]
	regardless of the accepted exchanges or the stopping round.
	
	Conditional on the final collection, uniform candidate
	selection includes unit $i$ with probability
	$\nu_i(\mathcal S^{(R)})/J$. Taking expectations,
	\[
	\Prb_p(i\in S)
	=
	\mathbb E\left[
	\frac{\nu_i(\mathcal S^{(R)})}{J}
	\right]
	=
	\mathbb E\left[
	\frac{\nu_i(\mathcal S^{(0)})}{J}
	\right]
	=\pi_i.
	\]
	Since every candidate contains $n$ distinct units, the
	selected sample has size $n$.
\end{proof}

\end{document}